\documentclass[twoside,reqno]{bjp}
\usepackage{graphicx}
\usepackage{amssymb,amsmath}
\usepackage{times}
\begin{document}

\sloppy \raggedbottom
\setcounter{page}{1}

%
%
%
%
\newcommand\rf[1]{(\ref{eq:#1})}
\newcommand\lab[1]{\label{eq:#1}}
\newcommand\nonu{\nonumber}
\newcommand\br{\begin{eqnarray}}
\newcommand\er{\end{eqnarray}}
\newcommand\be{\begin{equation}}
\newcommand\ee{\end{equation}}
\newcommand\eq{\!\!\!\! &=& \!\!\!\! }
\newcommand\foot[1]{\footnotemark\footnotetext{#1}}
\newcommand\lb{\lbrack}
\newcommand\rb{\rbrack}
\newcommand\llangle{\left\langle}
\newcommand\rrangle{\right\rangle}
\newcommand\blangle{\Bigl\langle}
\newcommand\brangle{\Bigr\rangle}
\newcommand\llb{\left\lbrack}
\newcommand\rrb{\right\rbrack}
\newcommand\Blb{\Bigl\lbrack}
\newcommand\Brb{\Bigr\rbrack}
\newcommand\lcurl{\left\{}
\newcommand\rcurl{\right\}}
\renewcommand\({\left(}
\renewcommand\){\right)}
\renewcommand\v{\vert}                     
\newcommand\bv{\bigm\vert}               
\newcommand\Bgv{\;\Bigg\vert}            
\newcommand\bgv{\bigg\vert}              
\newcommand\lskip{\vskip\baselineskip\vskip-\parskip\noindent}
\newcommand\mskp{\par\vskip 0.3cm \par\noindent}
\newcommand\sskp{\par\vskip 0.15cm \par\noindent}
\newcommand\bc{\begin{center}}
\newcommand\ec{\end{center}}
\newcommand\Lbf[1]{{\Large \textbf{{#1}}}}
\newcommand\lbf[1]{{\large \textbf{{#1}}}}




\newcommand\tr{\mathop{\mathrm tr}}                  
\newcommand\Tr{\mathop{\mathrm Tr}}                  
\newcommand\partder[2]{\frac{{\partial {#1}}}{{\partial {#2}}}}
\newcommand\partderd[2]{{{\partial^2 {#1}}\over{{\partial {#2}}^2}}}
\newcommand\partderh[3]{{{\partial^{#3} {#1}}\over{{\partial {#2}}^{#3}}}}
\newcommand\partderm[3]{{{\partial^2 {#1}}\over{\partial {#2} \partial{#3} }}}
\newcommand\partderM[6]{{{\partial^{#2} {#1}}\over{{\partial {#3}}^{#4}{\partial {#5}}^{#6} }}}          
\newcommand\funcder[2]{{{\delta {#1}}\over{\delta {#2}}}}
\newcommand\Bil[2]{\Bigl\langle {#1} \Bigg\vert {#2} \Bigr\rangle}  
\newcommand\bil[2]{\left\langle {#1} \bigg\vert {#2} \right\rangle} 
\newcommand\me[2]{\left\langle {#1}\right|\left. {#2} \right\rangle} 

\newcommand\sbr[2]{\left\lbrack\,{#1}\, ,\,{#2}\,\right\rbrack} 
\newcommand\Sbr[2]{\Bigl\lbrack\,{#1}\, ,\,{#2}\,\Bigr\rbrack}
\newcommand\Gbr[2]{\Bigl\lbrack\,{#1}\, ,\,{#2}\,\Bigr\} }
\newcommand\pbr[2]{\{\,{#1}\, ,\,{#2}\,\}}       
\newcommand\Pbr[2]{\Bigl\{ \,{#1}\, ,\,{#2}\,\Bigr\}}  
\newcommand\pbbr[2]{\lcurl\,{#1}\, ,\,{#2}\,\rcurl}




\renewcommand\a{\alpha}
\renewcommand\b{\beta}
\renewcommand\c{\chi}
\renewcommand\d{\delta}
\newcommand\D{\Delta}
\newcommand\eps{\epsilon}
\newcommand\vareps{\varepsilon}
\newcommand\g{\gamma}
\newcommand\G{\Gamma}
\newcommand\grad{\nabla}
\newcommand\h{\frac{1}{2}}
\renewcommand\k{\kappa}
\renewcommand\l{\lambda}
\renewcommand\L{\Lambda}
\newcommand\m{\mu}
\newcommand\n{\nu}
\newcommand\om{\omega}
\renewcommand\O{\Omega}
\newcommand\p{\phi}
\newcommand\vp{\varphi}
\renewcommand\P{\Phi}
\newcommand\pa{\partial}
\newcommand\tpa{{\tilde \partial}}
\newcommand\bpa{{\bar \partial}}
\newcommand\pr{\prime}
\newcommand\ra{\rightarrow}
\newcommand\lra{\longrightarrow}
\renewcommand\r{\rho}
\newcommand\s{\sigma}
\renewcommand\S{\Sigma}
\renewcommand\t{\tau}
\renewcommand\th{\theta}
\newcommand\bth{{\bar \theta}}
\newcommand\Th{\Theta}
\newcommand\z{\zeta}
\newcommand\ti{\tilde}
\newcommand\wti{\widetilde}
\newcommand\twomat[4]{\left(\begin{array}{cc}  
{#1} & {#2} \\ {#3} & {#4} \end{array} \right)}
\newcommand\threemat[9]{\left(\begin{array}{ccc}  
{#1} & {#2} & {#3}\\ {#4} & {#5} & {#6}\\
{#7} & {#8} & {#9} \end{array} \right)}


\newcommand\cA{{\mathcal A}}
\newcommand\cB{{\mathcal B}}
\newcommand\cC{{\mathcal C}}
\newcommand\cD{{\mathcal D}}
\newcommand\cE{{\mathcal E}}
\newcommand\cF{{\mathcal F}}
\newcommand\cG{{\mathcal G}}
\newcommand\cH{{\mathcal H}}
\newcommand\cI{{\mathcal I}}
\newcommand\cJ{{\mathcal J}}
\newcommand\cK{{\mathcal K}}
\newcommand\cL{{\mathcal L}}
\newcommand\cM{{\mathcal M}}
\newcommand\cN{{\mathcal N}}
\newcommand\cO{{\mathcal O}}
\newcommand\cP{{\mathcal P}}
\newcommand\cQ{{\mathcal Q}}
\newcommand\cR{{\mathcal R}}
\newcommand\cS{{\mathcal S}}
\newcommand\cT{{\mathcal T}}
\newcommand\cU{{\mathcal U}}
\newcommand\cV{{\mathcal V}}
\newcommand\cX{{\mathcal X}}
\newcommand\cW{{\mathcal W}}
\newcommand\cY{{\mathcal Y}}
\newcommand\cZ{{\mathcal Z}}

\newcommand{\nit}{\noindent}
\newcommand{\ct}[1]{\cite{#1}}
\newcommand{\bib}[1]{\bibitem{#1}}

\newcommand\PRL[3]{\textsl{Phys. Rev. Lett.} \textbf{#1} (#2) #3}
\newcommand\NPB[3]{\textsl{Nucl. Phys.} \textbf{B#1} (#2) #3}
\newcommand\NPBFS[4]{\textsl{Nucl. Phys.} \textbf{B#2} [FS#1] (#3) #4}
\newcommand\CMP[3]{\textsl{Commun. Math. Phys.} \textbf{#1} (#2) #3}
\newcommand\PRD[3]{\textsl{Phys. Rev.} \textbf{D#1} (#2) #3}
\newcommand\PLA[3]{\textsl{Phys. Lett.} \textbf{#1A} (#2) #3}
\newcommand\PLB[3]{\textsl{Phys. Lett.} \textbf{#1B} (#2) #3}
\newcommand\CQG[3]{\textsl{Class. Quantum Grav.} \textbf{#1} (#2) #3}
\newcommand\JMP[3]{\textsl{J. Math. Phys.} \textbf{#1} (#2) #3}
\newcommand\PTP[3]{\textsl{Prog. Theor. Phys.} \textbf{#1} (#2) #3}
\newcommand\SPTP[3]{\textsl{Suppl. Prog. Theor. Phys.} \textbf{#1} (#2) #3}
\newcommand\AoP[3]{\textsl{Ann. of Phys.} \textbf{#1} (#2) #3}
\newcommand\RMP[3]{\textsl{Rev. Mod. Phys.} \textbf{#1} (#2) #3}
\newcommand\PR[3]{\textsl{Phys. Reports} \textbf{#1} (#2) #3}
\newcommand\FAP[3]{\textsl{Funkt. Anal. Prilozheniya} \textbf{#1} (#2) #3}
\newcommand\FAaIA[3]{\textsl{Funct. Anal. Appl.} \textbf{#1} (#2) #3}
\newcommand\TAMS[3]{\textsl{Trans. Am. Math. Soc.} \textbf{#1} (#2) #3}
\newcommand\InvM[3]{\textsl{Invent. Math.} \textbf{#1} (#2) #3}
\newcommand\AdM[3]{\textsl{Advances in Math.} \textbf{#1} (#2) #3}
\newcommand\PNAS[3]{\textsl{Proc. Natl. Acad. Sci. USA} \textbf{#1} (#2) #3}
\newcommand\LMP[3]{\textsl{Letters in Math. Phys.} \textbf{#1} (#2) #3}
\newcommand\IJMPA[3]{\textsl{Int. J. Mod. Phys.} \textbf{A#1} (#2) #3}
\newcommand\IJMPD[3]{\textsl{Int. J. Mod. Phys.} \textbf{D#1} (#2) #3}
\newcommand\TMP[3]{\textsl{Theor. Math. Phys.} \textbf{#1} (#2) #3}
\newcommand\JPA[3]{\textsl{J. Physics} \textbf{A#1} (#2) #3}
\newcommand\JSM[3]{\textsl{J. Soviet Math.} \textbf{#1} (#2) #3}
\newcommand\MPLA[3]{\textsl{Mod. Phys. Lett.} \textbf{A#1} (#2) #3}
\newcommand\JETP[3]{\textsl{Sov. Phys. JETP} \textbf{#1} (#2) #3}
\newcommand\JETPL[3]{\textsl{ Sov. Phys. JETP Lett.} \textbf{#1} (#2) #3}
\newcommand\PHSA[3]{\textsl{Physica} \textbf{A#1} (#2) #3}
\newcommand\PHSD[3]{\textsl{Physica} \textbf{D#1} (#2) #3}
\newcommand\JPSJ[3]{\textsl{J. Phys. Soc. Jpn.} \textbf{#1} (#2) #3}
\newcommand\JGP[3]{\textsl{J. Geom. Phys.} \textbf{#1} (#2) #3}

\newcommand\Xdot{\stackrel{.}{X}}
\newcommand\xdot{\stackrel{.}{x}}
\newcommand\ydot{\stackrel{.}{y}}
\newcommand\yddot{\stackrel{..}{y}}
\newcommand\rdot{\stackrel{.}{r}}
\newcommand\rddot{\stackrel{..}{r}}
\newcommand\vpdot{\stackrel{.}{\varphi}}
\newcommand\vpddot{\stackrel{..}{\varphi}}
\newcommand\tdot{\stackrel{.}{t}}
\newcommand\zdot{\stackrel{.}{z}}
\newcommand\etadot{\stackrel{.}{\eta}}
\newcommand\udot{\stackrel{.}{u}}
\newcommand\vdot{\stackrel{.}{v}}
\newcommand\rhodot{\stackrel{.}{\rho}}
\newcommand\xdotdot{\stackrel{..}{x}}
\newcommand\ydotdot{\stackrel{..}{y}}



\title{Lightlike Membranes in Black Hole and  Wormhole Physics, and Cosmology
\thanks{Talk at Second Bulgarian National Congress in Physics, Sept. 2013}}

\begin{start}
\author{E.~Guendelman}{1}, \coauthor{A.~Kaganovich}{1},
\coauthor{E.~Nissimov}{2}, \coauthor{S.~Pacheva}{2}

\address{Department of Physics, Ben-Gurion Univ. of the Negev,
Beer-Sheva 84105, Israel}{1}

\address{Institute of Nuclear Research and Nuclear Energy,
Bulg. Acad. Sci., Sofia 1784, Bulgaria}{2}

\runningheads{E.~Guendelman, A.~Kaganovich, E.~Nissimov, S.~Pacheva}{Lightlike 
Membranes in Black Hole $\&$ Wormhole Physics $\&$ Cosmology}

\received{}

\begin{Abstract}
We shortly outline the principal results concerning the reparametrization-invariant
world-volume Lagrangian formulation of {\em lightlike} brane dynamics and
its impact as a source for gravity and (nonlinear) electromagnetism in black hole 
and wormhole physics.
\end{Abstract}


\end{start}

\section[]{Introduction}

Extended objects (strings and p-branes) are of primary importance for the
construction of self-consistent unified modern theory of fundamental forces in Nature
\ct{polchinski}.
In a series of recent papers of ours \ct{LL-main-1,Reg-BH,BR-kink} 
we have proposed for the first time 
in the literature a systematic world-volume Lagrangian description and studied in 
detail the physical properties of a new class of brane theories called 
{\em lightlike branes} (\textsl{LL-branes}), which are qualitatively distinct 
from the standard Nambu-Goto
type brane models which describe intrinsically {\em massive} world-volume modes.

As it is well known, \textsl{LL-branes} (also called \textsl{null-branes}) 
are of substantial interest in general relativity as they describe impulsive 
lightlike signals arising in various violent astrophysical events, \textsl{e.g.}, 
final explosion in cataclysmic processes such as supernovae and collision of 
neutron stars. \textsl{LL-branes} also play important
role in the description of various other physically important cosmological and 
astrophysical phenomena such as the ``membrane paradigm'' of black hole physics
and the thin-wall approach to domain walls coupled to gravity. For a detailed 
account, see \ct{barrabes-hogan}. More recently they became significant also in the
context of modern non-perturbative string theory \ct{nonperturb-string}.

Here we will shortly describe some of our principal results concerning the physics of
\textsl{LL-branes} and their implications in black hole and wormhole physics, 
and cosmology:

(a) Horizon ``straddling'' effect: the dynamics of \textsl{LL-branes} requires the 
bulk space-time geometry to possess one or more horizons, for instance, 
to be of black hole type, and it dictates that \textsl{LL-branes} automatically occupy 
(one of these) horizon(s).

(b) \textsl{LL-branes} are natural candidates for matter and charged sources of
``thin-shell'' traversable wormholes of various types (one- or multi-``throat''
``tube-like'', rotating etc.) \ct{LL-main-1}.

(c) \textsl{LL-branes} naturally produce regular black holes, i.e., black holes 
free of ``inside'' (below the inner horizon) physical space-time singularities
\ct{Reg-BH}.

(d) \textsl{LL-branes} trigger spontaneous compactification of space-time, 
as well as compactification/decompactification transitions \ct{BR-kink}.

(e) \textsl{LL-branes} are consistent matter sources for lightlike braneworlds 
\ct{Varna-11}.

(f) \textsl{LL-branes} produce new wormhole ``universes'' exhibiting 
{\em charge-hiding} and {\em charge-confining} effects \ct{hide-confine}, physically analogous 
to the quark confinement mechanism in quantum chromodynamics.

\section{Gravity and Nonlinear Gauge Fields Coupled to LL-Brane Sources}

In Refs.\ct{LL-main-1,Reg-BH,BR-kink,Varna-11,hide-confine} we proposed and 
extensively studied a manifestly reparametrization
invariant world-volume Lagrangian action of \textsl{LL-branes}:
\br
S_{\rm LL}\lb q\rb  = - \h \int d^{p+1}\s\, T b_0^{\frac{p-1}{2}}\sqrt{-\g}
\llb \g^{ab} {\bar g}_{ab} - b_0 (p-1)\rrb \; ,
\lab{LL-action+EM} \\
{\bar g}_{ab} \equiv 
g_{ab} - \frac{1}{T^2} (\pa_a u + q\cA_a)(\pa_b u  + q\cA_b) 
\quad , \quad \cA_a \equiv \pa_a X^\m A_\m \; .
\lab{ind-metric-ext-A}
\er
Here and below the following notations are used:
\begin{itemize}
\item
$\g_{ab}$ is the {\em intrinsic} world-volume Riemannian metric;
$g_{ab}=\pa_a X^{\m} G_{\m\n}(X) \pa_b X^{\n}$ is the {\em induced} metric on the 
world-volume, which becomes {\em singular} on-shell (manifestation of the lightlike 
nature); $b_0$ is world-volume ``cosmological constant'';
\item
$X^\m (\s)$ are the $p$-brane embedding coordinates in the
$D$-dimensional bulk space-time with Riemannian metric
$G_{\m\n}(x)$ ($\m,\n = 0,1,\ldots ,D-1$); 
$(\s)\equiv \(\s^0 \equiv \t,\s^i\)$ with $i=1,\ldots ,p$;
$\pa_a \equiv \partder{}{\s^a}$.
\item
$u$ is auxiliary world-volume scalar field defining the lightlike direction
of the induced metric;
\item
$T$ is {\em dynamical (variable)} brane tension;
\item
$q$ -- the coupling to bulk spacetime gauge field $\cA_\m$ is
\textsl{LL-brane} surface charge density.
\end{itemize}

The on-shell singularity, \textsl{i.e.}, the lightlike property of the induced metric
$g_{ab}$, directly follows from the equations of motion resulting from
\rf{LL-action+EM}:
\be
g_{ab} \({\bar g}^{bc}(\pa_c u  + q\cA_c)\) = 0 \; .
\lab{on-shell-singular-A}
\ee

Now, the full action of gravity and (nonlinear) gauge fields interacting
self-consistently with \textsl{LL-branes} reads (we specialize to $D=4$
space-time dimensions and use units with the Newton constant $G_N=1$):
\be
S = \int d^4 x \sqrt{-G} \Bigl\lb \frac{R(G) - 2\L_0}{16\pi} + L(F^2)\Bigr\rb 
+ \sum_{k=1}^N S_{\mathrm{LL}}\lb q^{(k)}\rb \; ,
\lab{gravity+GG+LL}
\ee
where the superscript $(k)$ indicates the $k$-th \textsl{LL-brane}. Here 
$R(G)= G^{\m\n}R_{\m\n}$ and $R_{\m\n}$ denote the Riemannian scalar curvature 
and the Ricci tensor of the bulk space-time geometry. $L(F^2)$ is the Lagrangian 
of a remarkable non-standard nonlinear electrodynamics containing 
\textsl{square root} of ordinary Maxwell Lagrangian \ct{GG}:
\be
L(F^2) = - \frac{1}{4} F^2 - \frac{f_0}{2} \sqrt{- F^2} \quad ,\quad
F^2 \equiv F_{\m\k} F_{\n\l} G^{\m\n} G^{\k\l} \; .
\lab{GG}
\ee
This is an explicit realization of `t Hooft's proposal (in flat space-time) for 
\textsl{infrared charge confinement} \ct{thooft} (see also next talk \ct{BJP-svet} at this congress).

\section{Charge-Confinement via ``Tube-Like'' Wormhole}

The general scheme to construct ``lightlike thin-shell'' wormholes of 
static ``spherically-symmetric'' type (in Eddington-Finkelstein coordinates
$dt=dv-\frac{d\eta}{A(\eta)}$ and ``radial''-like coordinate
$\eta \in (-\infty,+\infty)$):
\br
ds^2 = - A(\eta) dv^2 + 2dv d\eta + C(\eta) h_{ij}(\th) d\th^i d\th^j \quad ,\quad
F_{v\eta} = F_{v\eta} (\eta)\; , 
\lab{static-spherical-EF} \\
-\infty < \eta < \infty \quad, \;\; A(\eta^{(k)}_0) = 0 \;\; 
\mathrm{for} \;\; \eta^{(1)}_0 <\ldots<\eta^{(N)}_0 
\lab{common-horizons}
\er
is as follows (cf. Section 5 in Ref.\ct{hide-confine}):

(1) Take ``vacuum'' solutions of Einstein and (nonlinear) Maxwell equations
resulting from \rf{gravity+GG+LL} (\textsl{i.e.}, without the delta-function 
\textsl{LL-brane} contributions) in each space-time region (separate ``universe'') 
given by $\bigl(-\infty\! <\!\eta\!<\!\eta^{(1)}_0\bigr),\ldots,$
$\bigl(\eta^{(N)}_0 \!<\!\eta \!<\!\infty\bigr)$ with common horizon(s) at 
$\eta=\eta^{(k)}_0$ ($k=1,\ldots ,N$).

(2) Each $k$-th \textsl{LL-brane} automatically locates itself on the
horizon at $\eta=\eta^{(k)}_0$ -- intrinsic property of \textsl{LL-brane} dynamics 
defined by the action \rf{LL-action+EM}. 

(3) Match discontinuities of the derivatives of the metric and
the gauge field strength  across each horizon at $\eta=\eta^{(k)}_0$ using the 
explicit expressions for the \textsl{LL-brane} stress-energy tensor and charge 
current density systematically derived from the action \rf{gravity+GG+LL} with
\rf{LL-action+EM}.

Let us now consider the gravity/nonlinear-gauge-field system coupled to two
oppositely charged \textsl{LL-branes}, \textsl{i.e.}, $N=2$ and $q_1 = -
q_2 \equiv q$ in \rf{gravity+GG+LL}. We obtain a particularly interesting
``two-throat'' wormhole-type solution exhibiting a QCD-like charge
confinement effect. The total space-time manifold consists of three
``universes'' with different geometry glued together at their common
horizons occupied by the two oppositely charged \textsl{LL-branes}: 

(i) ``Left-most'' non-compact ``universe'' comprising the exterior
region of a new kind of {\em non-standard} Schwarzschild-de-Sitter-type black hole, 
with additional {\em constant vacuum radial electric field} $\vec{E}_{\rm vac}$,
beyond the Schwarzschild-type horizon $r_0$ for the ``radial-like'' 
$\eta$-coordinate interval $-\infty < \eta < -\eta_0 
\equiv - \Bigl\lb 4\pi\(\sqrt{2}f_0|\vec{E}| - \vec{E}^2\) + \L_0 \Bigr\rb^{-\h}$,
where (using notations as in \rf{static-spherical-EF}):
\br
A(\eta) = 1 - \frac{2m}{r_0 - \eta_0 - \eta} 
- \frac{\L_{\rm eff}}{3} (r_0 - \eta_0 - \eta)^2 \; ,
\lab{SdS-left-most-1}\\
C(\eta) = (r_0 - \eta_0 - \eta)^2 \;\; ,\;\;
|F_{v\eta}(\eta)| \equiv |\vec{E}_{\rm vac}| = \frac{f_0}{\sqrt{2}} <|\vec{E}| \; .
\lab{SdS-left-most-2}
\er
Here $\vec{E}$ is the constant electric field in the ``middle'' ``tube-like''
``universe'' (ii) (Eq.\rf{LCBR-middle-2} below);
$\L_{\rm eff}\equiv \L_0 + 2\pi f_0^2$ in \rf{SdS-left-most-1}
is {\em dynamically generated/shifted} cosmological constant, which is non-vanishing
even in the absence of the ``bare'' cosmological constant $\L_0$. Let us stress
that {\em constant vacuum radial electric fields} such as in \rf{SdS-left-most-2}
{\em do not} exist as solutions of ordinary Maxwell electrodynamics on
generic non-compact space-times -- the former are due exclusively to the nonlinear
``square-root'' term in \rf{GG}.

(ii) ``Middle'' ``tube-like'' ``universe'' of
Levi-Civita-Bertotti-Robinson type \ct{LC-BR} with geometry $dS_2 \times S^2$ 
($dS_2$ denotes two-dimensional de Sitter space; $S^2$ -- sphere with
constant radius $r_0$), comprising the finite extent 
(w.r.t. $\eta$-coordinate) region between the two horizons of $dS_2$ at 
$\eta = \pm \eta_0$ occupied by the two \textsl{LL-branes} with charges $\pm q$:
\be
-\eta_0 < \eta < \eta_0 
\equiv \Bigl\lb 4\pi\(\sqrt{2}f_0|\vec{E}| - \vec{E}^2\) + \L_0 \Bigr\rb^{-\h} \; ,
\lab{middle-interval}
\ee
where the metric coefficients and electric field are:
\br
A(\eta) = 
1 - \Bigl\lb 4\pi\(\sqrt{2}f_0|\vec{E}| - \vec{E}^2\) + \L_0 \Bigr\rb\,\eta^2
\;\; ,\;\; A(\pm \eta_0) = 0 \; ,
\lab{LCBR-middle-1} \\
C(\eta) = r_0^2 = \frac{1}{4\pi \vec{E}^2 + \L_0} \;\; ,\;\;
|\vec{E}| = |q| + \frac{f_0}{\sqrt{2}} = {\rm const} \; .
\lab{LCBR-middle-2}
\er

(iii) ``Right-most'' non-compact ``universe'' of the same type as (i) above
for the ``radial-like'' $\eta$-coordinate interval ~$\eta_0 < \eta < \infty$ 
($\eta_0$ as in \rf{middle-interval}). Its metric is given by 
Eq.\rf{SdS-left-most-1} upon changing $-\eta \to \eta$ and the elecric
field is the same as in \rf{SdS-left-most-2}.

The equations for the electric field (second relations in \rf{SdS-left-most-2} and 
\rf{LCBR-middle-2}) have profound consequences:

\begin{itemize}
\item
The ``left-most'' and ``right-most'' non-compact ``universes'' are
two identical copies of the {\em electrically neutral} exterior region of
Schwarzschild-de-Sitter black hole beyond the Schwarzschild horizon. They both
carry a constant vacuum radial electric field with magnitude 
$|\vec{E}|=\frac{f_0}{\sqrt{2}}$ pointing inbound/outbound w.r.t. pertinent
horizon. The corresponding electric displacement field 
$\vec{D}=\Bigl( 1 - \frac{f_0}{\sqrt{2}|\vec{E}|}\Bigr)\,\vec{E}=0$, 
so there is {\em no} electric flux there.
\item
The whole electric flux produced by the two charged \textsl{LL-branes} with 
opposite charges $\pm q$ at the boundaries of the above non-compact ``universes''
is {\em confined} within the finite-extent ``tube-like'' middle ``universe'' of 
Levi-Civitta-Robinson-Bertotti type
with geometry $dS_2 \times S^2$, where the constant electric field is
$|\vec{E}|=\frac{f_0}{\sqrt{2}} + |q|$ with associated non-zero electric 
displacement field $|\vec{D}|= |q|$ . This is {\em QCD-like confinement}.
\end{itemize}

The {\em charge-confining} wormhole geometry is visualized on Fig.1.
\begin{figure}
\begin{center}
\includegraphics[scale=0.7,angle=270,keepaspectratio=true]{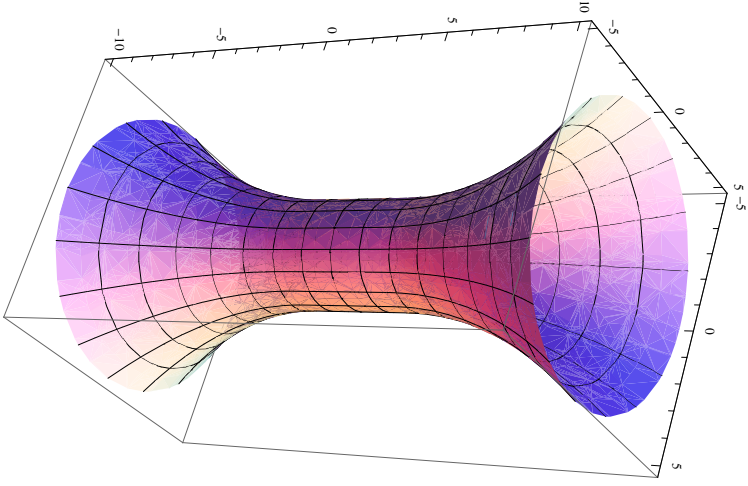}
\caption{Shape of $t=const$ and $\th=\frac{\pi}{2}$ slice of
charge-confining wormhole geometry. The whole electric flux is confined
within the middle cylindric tube.}
\end{center}
\end{figure}

To conclude let us emphasize that the existence of charge-confining
``thin-shell'' wormholes is entirely due to the combined effect of the exceptional 
properties of \textsl{LL-brane} dynamics and the ``square-root'' nonlinear electrodynamics.



\section*{Acknowledgments}
We gratefully acknowledge support of our collaboration through the academic exchange 
agreement between the Ben-Gurion University and the Bulgarian Academy of Sciences.
S.P. has received partial support from COST action MP-1210.


\end{document}